\documentclass[aps,prd,nofootinbib,floatfix,showpacs,preprintnumbers]{revtex4}

\usepackage{graphicx}
\newcommand{\sfig}[2]{
\includegraphics[width=#2]{#1}
        }
\newcommand{\Sfig}[2]{
    \begin{figure}[thbp]
    \sfig{#1.eps}{.6\columnwidth}
    \caption{{\small #2}}
    \label{fig:#1}
    \end{figure}
}

\newcommand{\rf}[1]{\ref{fig:#1}}

\newcommand\bw{(B\omega)^{\rm meas}}
\newcommand\sww{S^{\omega\omega}}
\newcommand\sbb{S^{BB}}
\newcommand\nww{1}
\newcommand\nbb{1}
\newcommand\varcw{\Delta^2}

\def\lsim{\mathrel{\raise.3ex\hbox{$<$\kern-.75em\lower1ex\hbox{$\sim$}}}}
\def\gsim{\mathrel{\raise.3ex\hbox{$>$\kern-.75em\lower1ex\hbox{$\sim$}}}}

\def\cmm2{{\,\rm cm^{-2}}}
\def\cm2{{\,{\rm cm}^2}}
\def\cmm3{{\,{\rm cm}^{-3}}}
\def\gcmm3{{\,{\rm g\,cm^{-3}}}}

\def\fun#1#2{\lower3.6pt\vbox{\baselineskip0pt\lineskip.9pt
  \ialign{$\mathsurround=0pt#1\hfil##\hfil$\crcr#2\crcr\sim\crcr}}}

\def\be{\begin{equation}}
\def\ee{\end{equation}}
\def\bea{\begin{eqnarray}}
\def\eea{\end{eqnarray}}
\newcommand{\vs}{\nonumber\\}

\newcommand{\ec}[1]{Eq.~(\ref{eq:#1})}
\newcommand{\Ec}[1]{(\ref{eq:#1})}
\newcommand{\eql}[1]{\label{eq:#1}}

\begin{document}

\preprint{FERMILAB-PUB-10-020-A-PPD-T}

%\vspace{.2in}
\title{Cross-Correlating Probes of Primordial Gravitational Waves}

%\vspace{.2in}
\author{Scott Dodelson$^{1,2,3}$}
%\vspace{.2in}

\affiliation{$^1$Center for Particle Astrophysics, Fermi National
Accelerator Laboratory, Batavia, IL~~60510-0500, USA}
\affiliation{$^2$Department of Astronomy \& Astrophysics, The
University of Chicago, Chicago, IL~~60637-1433, USA}
\affiliation{$^3$Kavli Institute for Cosmological Physics, Chicago, IL~~60637-1433, USA}
\date{\today}
%\smallskip
\begin{abstract}
One of the most promising ways of detecting primordial gravitational waves generated during inflation is to observe B-modes
of polarization, generated by Thomson scattering after reionization, in the cosmic microwave background (CMB). Large scale foregrounds though are expected to be a major systematic issue, so -- in the event of a tentative detection -- an independent confirmation of large scale gravitational waves would be most welcome. Previous authors have suggested searching for the analogous mode of cosmic shear in weak lensing surveys but have shown that the signal to noise of this mode is marginal at best. This argument is reconsidered here, accounting for the cross-correlations of the polarization and lensing B-modes. A lensing survey can potentially strengthen the argument for a detection of primordial gravitational waves, although it is unlikely to help constrain the amplitude of the signal.
\end{abstract}
\maketitle

\section{introduction}

Many models of inflation predict that gravitational waves were produced along with density perturbations in the early universe~\cite{1982PhLB..115..189R,1983PhLB..125..445F,1984NuPhB.244..541A,1988PhRvD..37.2078A,1993PhRvL..70.2371G}. While the density perturbations have been detected and mapped out very precisely, the gravitational waves have yet to be detected. Perhaps our best hope is to observe the unique pattern of polarization -- so-called B-modes -- that gravitational waves produce when photons in a warped radiation field scatter off free electrons~\cite{Seljak:1996gy,Kamionkowski:1996zd,Baumann:2008aq,Dodelson:2009kq}. One exciting feature of this signal is that it was produced at two distinct epochs: first, during recombination when the scattering rate starts to tail off and photons can travel longer distances, and, then, after reionization at much later times. These two epochs are imprinted in the angular spectrum on large scales (multipoles $l\sim 2-10$) and small scales ($\sim50-100$). Hence primordial gravitational waves leave the distinct signature of a double-humped peak in the power spectrum of polarization B-modes. 

Detection, however, is far from guaranteed. Among the hurdles that must be overcome are: (i) unknown amplitude of the signal~\cite{Lyth:1996im}, (ii) high sensitivity~\cite{Bock:2006yf} needed to detect even the largest signal possible, and (iii) foreground contamination~\cite{Dunkley:2008am}. Even if the first two of these are overcome, and a B-mode detection is claimed, distinguishing a putative signal from foregrounds is likely to remain a significant problem. Here, I explore one possible way to test a future B-mode detection against foreground contamination: supplementing the polarization B-mode signal with an all-sky, deep weak lensing signal. 

The idea is simple: gravitational waves also leave an imprint on the cosmic distortion tensor that governs the propagation of light over long distances, and hence the observed shapes of galaxies. Density perturbations in the distortion tensor produce {\it scalar shear}, which was first detected in 2000~\cite{vanWaerbeke:2000rm,Wittman:2000tc,Kaiser:2000if,Bacon:2000sy} and has since been measured over larger scales by many experiments (see, e.g., \cite{Hoekstra:2008db} for a review), but gravitational waves produce a different type of distortion~\cite{Stebbins:1996wx}, sometimes called the pseudo-scalar shear, or the {\it curl} mode of shear, the exact analogue of the B-mode of the polarization field. A number of studies~\cite{Dodelson:2003bv,Sarkar:2008ii} have shown that the power spectrum of this mode, even in the most optimistic scenarios (large amplitude, all-sky survey with low shape noise), will be difficult to detect except perhaps at the lowest multipoles. Here I point out that the situation is not quite as grim as these studies suggest. First, even a moderate signal to noise independent detection in an arena with completely different systematics would significantly increase our confidence in a putative CMB detection. Further, the cross-correlation between the CMB and lensing B-modes is non-zero and this helps the search for a detection.

The main purpose of this paper is to calculate the extent to which this is true, the extent to which the same modes which produce the CMB B-mode reionization signal are responsible for the B-mode in a lensing experiment. This is an interesting academic question in its own right: two very different phenomena owe their existence to the same spectrum of gravitational waves, and it is interesting to quantify how correlated they are. 
Sections II and III present the formula for the auto-correlations of the B-modes in these two different arenas. Since I am interested only in the large scale feature, the focus in \S{II} is on the reionization signal, and, as far as I can tell, the final formula for this signal in Eqs. \Ec{clbint} and \Ec{clbfin}
has not been presented elsewhere. In \S{IV}, I calculate the cross-correlation between the two and show that it is non-negligible. Finally, \S{V} addresses the more general question of how much additional information can be gleaned from a lensing survey, taking into account the cross-correlation.
Some of the details of the calculations are relegated to appendices.

There is one scenario for which the probe introduced here would be of tremendous utility. Over a decade ago, cosmologists studied gravitational wave production in open inflation models~\cite{GarciaBellido:1997hy,Tanaka:1997kq,Linde:1999wv,Hertog:1999kg,Hawking:2000ee} and concluded that the amplitude would be boosted near the horizon. This would lead to a much different spectrum than the double humped peak which modulates the scale-invariant standard inflation spectrum. Recently, there have been suggestions~\cite{Freivogel:2005vv,Susskind:2007pv} that inflation was preceeded by an earlier epoch of vacuum domination such that our Universe was produced via a Coleman-De Luccia tunneling process~\cite{Coleman:1980aw} and is therefore open, albeit with a small value of the curvature density. Even with the small curvature expected in our Universe, though, the large scale gravitational wave spectrum might be different than the simple scale-invariant one generally assumed. 
%even a slightly open universe would Susskind and collaborators have recently explored a similar idea with a twist: an ancestor epoch of false vacuum domination preceded standard inflation~\cite{Freivogel:2005vv,Susskind:2007pv}. This pre-inflation stage likely occurred at a very high energy scale and therefore the gravitational wave background amplitude during it was likely very 
%large. These pre-primordial gravitational waves may have leaked into the bubble associated with our universe, but only on the largest scales. In this scenario, too, the B-mode spectrum would differ significantly from the standard double-humped prediction. Rather, one would find relatively large amplitudes at $l=2,3$, rapidly falling on smaller scales. 
A test -- such as the B-mode of cosmic shear -- sensitive to only the largest moments seems tailor-made to probe this class of models.

\section{B-modes of Polarization}

The CMB radiation is described by a $2\times2$ intensity tensor (see, e.g., Ref.~\cite{1997PhRvD..55.1830Z,Cabella:2004mk}), the traceless, symmetric part of which consists of two fields $Q(\hat n)$ and $U(\hat n)$.
%\begin{equation}
%I_{ab} =2\left( \matrix{ I + Q & U\cr U & I-Q}\right).\eql{defqu}
%\end{equation}
It is useful to decompose $Q$ and $U$ into E/B modes, where the B-mode multipole moments are given by
\begin{equation}
B_{lm} = \frac{i}{2}\int d^2n \left[
\big({}_2Y^*_{lm}(\hat n)\big) \big(Q+iU)(\hat n) -
\big({}_{-2}Y^*_{lm}(\hat n)\big) \big(Q-iU)(\hat n) 
\right]\eql{defB}
\end{equation}
where the ${}_{\pm2}Y_{lm}$ are the spin-2 spherical harmonics.
The polarization fields $Q(\hat n)$ and $U(\hat n)$ are induced by Thomson scattering in the presence of a quadrupole distribution, so
the polarization produced after reionization is
\begin{equation}
\left(Q\pm iU\right)(\hat n) = \frac{3}{5\sqrt{6}} \int_{0}^{D_{\rm reion}} dD\, \dot\tau(\eta)
\sum_{m''=-2}^2\big({}_{\pm2}Y^*_{lm''}(\hat n)\big) \Theta_{2m''}(\vec x=D\hat n;\eta).\eql{qpmu}
\end{equation}
where $\eta$ is the conformal time associated with the comoving distance $D$ (in the flat universe we will assume throughout, $\eta=\eta_0-D$).
That is, one integrates along the line of sight out to a spherical shell a distance $D_{\rm reion}$ away (see Fig. \rf{circ}, adapted from Ref.~\cite{Dvorkin:2007jp}), weighting by the scattering rate $\dot\tau=n_e\sigma_T a$ (where $n_e$ is the free electron number density, $\sigma_T$ the Thomson cross-section, and $a$ the scale factor), and 
the different components of the quadrupole $\Theta_{2m}$. 
Eq.~\Ec{qpmu} relates the polarization field produced after reionization to the quadrupole of the radiation field, no matter what the origin of that quadrupole. The quadrupole produced by gravitational waves (GW)
is of course linearly related to the tensor perturbation $h_{ij}(\vec x,\eta)$ with~(e.g., Ref. \cite{2009AIPC.1132...86C})
\begin{equation}
\Theta_{2m}(\vec x,\eta) = \int d^2 n'\, Y^*_{2m}(\hat n') 
\int_{0}^{\Delta D} d(\Delta D') \left[ \frac{-1}{2} \hat n'^i \hat n'^j\dot h_{ij}(\vec x+\vec x';\eta')\right]\eql{quad}
\end{equation}
where $\Delta D \equiv D_*-D$ is the distance between the point of interest and the surface of last scattering, $\dot h \equiv \partial h/\partial \eta'$, and 
$\vec x'\equiv \Delta D'\hat n'$ identifies a point within a sphere of radius $\Delta D$ centered at $\vec x$ (see Fig. \rf{circ}). 
To be clear, the argument $\eta'$ in the tensor perturbation denotes the conformal time when the GW (which travels at the speed of light) was at position $\vec x + \vec x'$. That is, it was a distance $\Delta D'$ away from the point $[\vec x=(D\hat n)]$. This corresponds to conformal time $\eta'=\eta-\Delta D'=\eta_0-(D+\Delta D')$. 

\Sfig{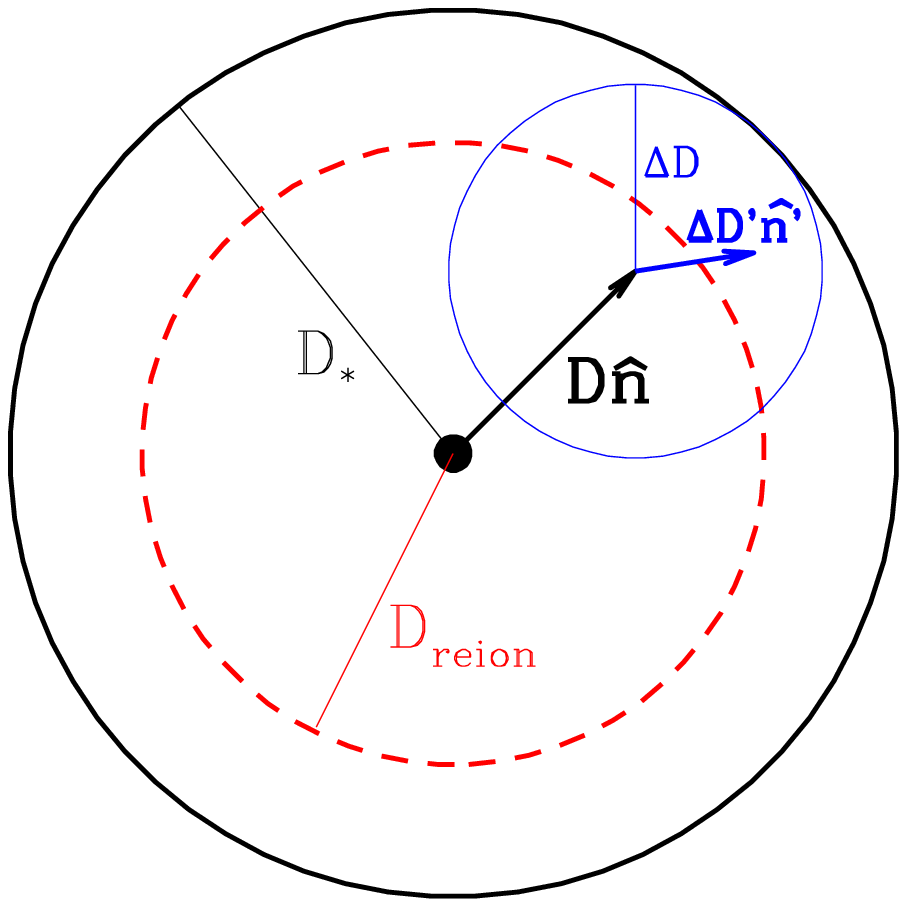}{The double integral which determines the B-mode signal due to reionization. The outer integral (\ec{qpmu}) is along the line of sight 
$D\hat n$ out to the time/distance of reionization, $D_{\rm reion}$. The quadrupole in the integrand is determined by an integral along $\Delta D'\hat n'$ 
(\ec{quad}) out to the surface of last scattering (a distance $\Delta D$ away) to capture the contribution of the gravitational waves.}

The tensor field is the sum of two independent modes in Fourier space:
\begin{equation}
h_{ij}(\vec x;\eta) = \int \frac{d^3k}{(2\pi)^3} e^{i\vec k\cdot\vec x} 
T(k,\eta) \sum_{\alpha=+,\times} \tilde h^{(\alpha)}(\vec k) \epsilon^{(\alpha)}_{ij}(\vec k)
\end{equation}
where $T(k,\eta)$ captures the evolution of the GW when they enter the horizon. In a matter dominated universe, $T(k,\eta)=3j_1(k\eta)/(k\eta)$, but we use the exact solution to the transfer function here which differs slightly at late times due to dark energy domination.
The orientation of the two modes depends on the direction of the $\hat k$ vector. If $\hat k$ is chosen to lie along the $z$-axis,
then
\begin{equation}
\epsilon^{(+)}_{ij}= \left(\matrix{1 &0 &0\cr 0 & -1 &0\cr 0 &0&0}\right)
\qquad\qquad
\epsilon^{(\times)}_{ij}= \left(\matrix{0 &1 &0\cr 1 & 0 &0\cr 0 &0&0}\right)
.
\end{equation}
More generally, they are transverse, traceless matrices normalized so that Tr[$e^{(\alpha)} e^{(\beta)}] =2\delta_{\alpha\beta}$.

Armed with these results, we can write
\begin{equation}
B_{lm} = \int\frac{d^3k}{(2\pi)^3} \sum_{\alpha=+,\times} \tilde h^{(\alpha)}(\vec k) T^{P,(\alpha)}_{lm}(\vec k)\eql{blm}
\end{equation}
where the transfer function for the B-modes in polarization is
\begin{equation}
T^{P,(\alpha)}_{lm}(\vec k) = -i\sqrt{\frac{3}{800}} 
\int_{0}^{D_{\rm reion}} dD\, \dot\tau \sum_{m''=-2}^2 I_{lmm''}(\vec k,D) \int_{0}^{\Delta D} d(\Delta D') \, \dot T(k,\eta')
 J^{(\alpha)}_{m''}(\vec k,\Delta D')  ;\eql{tran}
\end{equation} 
\begin{equation}
J^{(\alpha)}_{m''}(\vec k,\Delta D') \equiv \epsilon_{ij}^{(\alpha)}(\vec k) \int d^2n'\, Y^*_{2m''}(\hat n') \hat n'^i \hat n'^j
e^{i\vec k\cdot \hat n' \Delta D'} \eql{defj};
\end{equation}
and 
\begin{equation}
I_{lmm''}(\vec k,D) \equiv \int d^2n\,  e^{i\vec k\cdot \hat n D} \left[ \big({}_{2}Y^*_{lm}(\hat n)\big)
\big({}_{2}Y_{2m''}(\hat n)\big)-
\big({}_{-2}Y^*_{lm}(\hat n)\big)\big({}_{-2}Y_{2m''}(\hat n)\big)\right].\eql{defi}
\end{equation}

The transfer function is difficult to compute for general $\vec k$ but simplifies considerably when $\vec k$ lies along the $\hat z$-axis. 
Fortunately, the auto- and cross-spectra that can be observed are rotationally invariant so in the end we will need only this simple case. 
The calculation is presented in Appendix~\ref{app:pol} with the result that
\begin{eqnarray}
T^{P,(+)}_{lm}(k\hat z) &=& i^{l+1}\sqrt{\frac{9\pi}{4(2l+1)}}  \left[\delta_{m,2}-\delta_{m,-2}\right]
\int_{0}^{D_{\rm reion}} dD\, \dot\tau  \left[ (l+2) j_{l-1}(kD) -(l-1) j_{l+1}(kD) \right] \nonumber\\
&&\times \int_{0}^{\Delta D} d(\Delta D') \, \dot T(k,\eta')
\frac{j_2(k\Delta D')}{(k\Delta D')^2} .\eql{tranfin}
\end{eqnarray} 
The $m=2$ component is plotted for the lowest moments in Fig.~\rf{tlbb} for $z_{\rm reion}=10$.

\Sfig{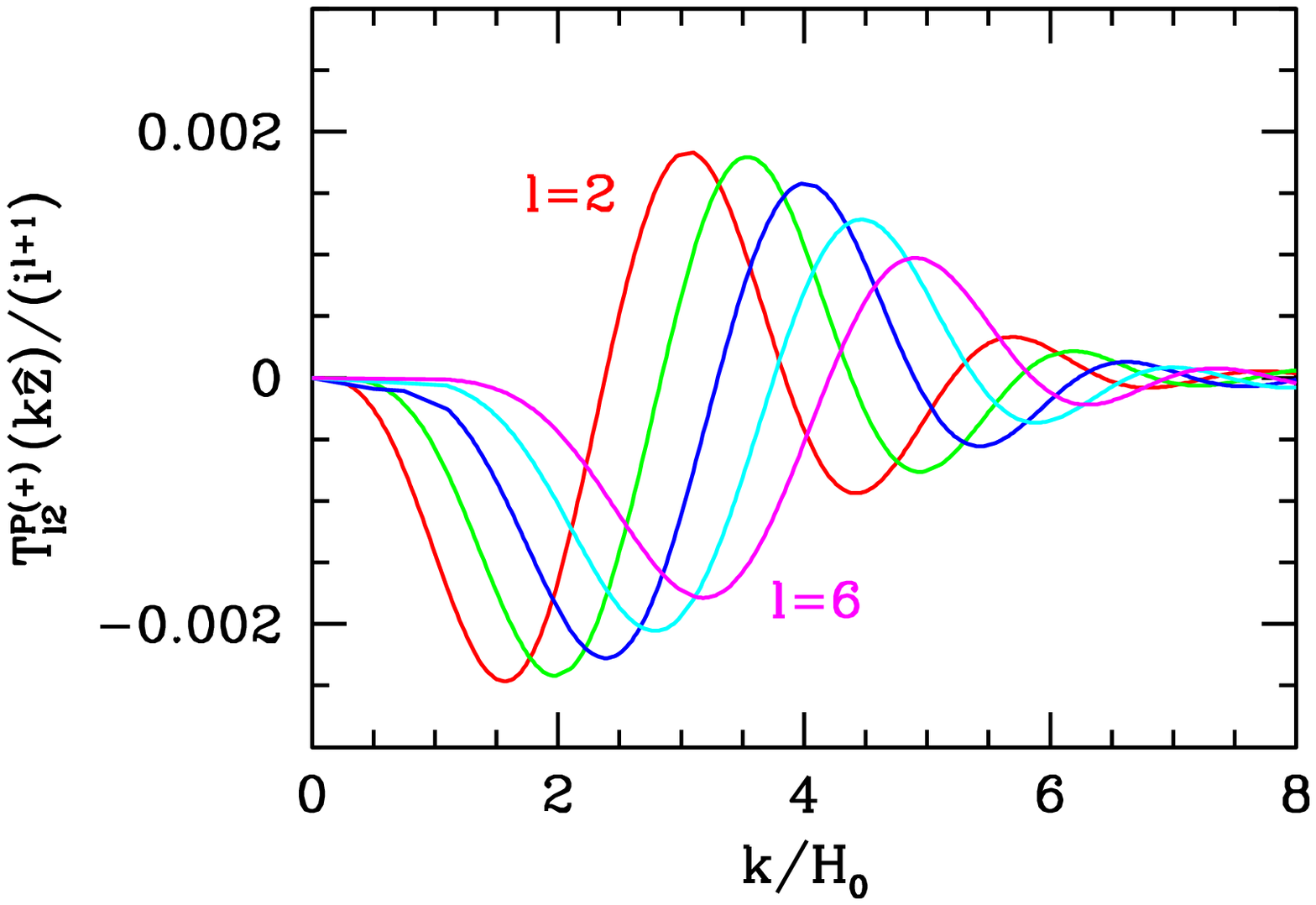}{The $m=2$ moment of the B-mode transfer function in polarization for $l=2-6$ as a function of wavenumber $k$. Here reionization is assumed to occur instantaneously at $z=10$.}

We will use \ec{tranfin} to compute the cross-spectra with lensing. But, as long as we have it, we can first use it to derive a simple formula for the auto-spectrum of polarization B-modes due to reionization. Of course this spectrum is by now a standard feature of freely available codes which compute temperature and polarization two-point functions~\cite{cmbfast,camb}, but it is nice to capture the physics in a simple semi-analytic formula.  The amplitude of the GW mode is drawn from a Gaussian distribution with the same power spectrum:
\begin{equation}
\langle \tilde h^{(\alpha)}(\vec k) \tilde h^{(\beta)\dagger}(\vec k')\rangle
=(2\pi)^3 \delta_{\alpha\beta}\delta^3(\vec k-\vec k') P_h(k).\eql{ph}
\end{equation} 
In standard slow roll inflation, the power spectrum is nearly scale invariant: $k^3 P_h(k)/(2\pi^2) =(4/\pi) (H_I/m_{\rm pl})^2$, where $H_I$ is the expansion rate during inflation. Squaring \ec{blm} and taking the expectation value using \ec{ph} then leads to the polarization B-mode spectrum
\begin{eqnarray}
C_l^{PP} &=& \frac{1}{2l+1}\sum_{m=-l}^l \langle \big\vert B_{lm} \big\vert^2 \rangle
\nonumber\\
&=& \frac{1}{2l+1}\int\frac{d^3k}{(2\pi)^3} P_h(k) W^{PP}_l(k),\eql{clbint}
\end{eqnarray}
where
\begin{eqnarray}
W^{PP}_l(k) &\equiv& \sum_{\alpha}\sum_{m=-l}^l\left\vert T_{lm}^{P,(\alpha)}(\vec k)\right\vert^2\vs 
&=&
\frac{9\pi}{2l+1} 
 \Bigg\vert\int_{0}^{D_{\rm reion}} dD\, \dot\tau\left(\eta_0-D\right) \vs &&\times \left[ (l+2) j_{l-1}(kD) -(l-1) j_{l+1}(kD) \right] \int_{0}^{\Delta D} d(\Delta D') \, \dot T(k,\eta')
\frac{j_2(k\Delta D')}{(k\Delta D')^2}\Bigg\vert^2.
\eql{clbfin}
\end{eqnarray}

\section{B-mode of Cosmic Shear}

The deformation tensor which describes propagation of light through the inhomogeneous universe can also be written as a $2\times2$ matrix, and its traceless, symmetric part can also be characterized by two fields: the two components of shear $\gamma_1$ and $\gamma_2$ replacing $Q$ and $U$. Thus, the moments of the B-mode of cosmic shear can be defined just as in \ec{defB}. There is a difference between the full deformation matrix and the full intensity matrix.
In addition to $Q$ and $U$, the polarization matrix contains a piece describing the intensity $I$ and circular polarization $V$. Neither $I$ nor $V$ is related to $Q$ and $U$ (in cosmology, $I$ of course describes the CMB temperature anisotropies while $V$ is thought to vanish because Thomson scattering induces no circular polarization). The analogous quantities in the lensing deformation tensor are the convergence $\kappa$ and the rotation $\omega$. These {\it are} related to the two components of the shear, with $\kappa$ equal to the $E$-mode and $\omega$ to the $B$-mode. Stebbins~\cite{Stebbins:1996wx} first examined the structure of this matrix and derived a number of useful relations among its components. Starting from the equivalent of \ec{defB} (with $\gamma_1$ and $\gamma_2$ replacing $Q$ and $U$), one arrives at a simple expression for the $B$-mode moments in terms of the rotation field
\begin{equation}
B_{lm} = \int d^2n Y_{lm}^*(\hat n) \omega(\hat n).
\end{equation}
The rotation receives no contributions from scalar perturbations, so it is non-zero only if tensor modes are present (at first order; see Ref.~\cite{Sarkar:2008ii} for second order scalar effects). Explicitly,
the moments of the B-mode of the shear field due to GW's are~\cite{Dodelson:2003bv}
\begin{equation}
B_{lm} = \int \frac{d^3k}{(2\pi)^3} \sum_{\alpha=+,\times} h^{(\alpha)}(\vec k) T^{L,(\alpha)}_{lm}(\vec k)
\end{equation}
with the lensing transfer function defined as
\begin{equation}
T^{L,(\alpha)}_{lm}(\vec k) \equiv \frac{1}{2} \int d^2n Y^*_{lm}(\hat n) \epsilon_{ijk} \hat n^i \hat n^l  \epsilon^{(\alpha)}_{kl}(\hat k) k_j
\int_0^{D_s} dD \, e^{i\vec k\cdot \hat n D} T(k,\eta_0-D) 
\end{equation}
where here $\epsilon_{ijk}$ is the 3D Levi-Civita symbol as opposed to the polarization tensor $\epsilon^{(\alpha)}_{kl}(\hat k)$. The integral here is out to the source galaxies, all assumed to be at distance $D_s$.

Again the integrals over angles simplify considerably when $\hat k$ is chosen to lie along the $z$-axis. Appendix~\ref{applen} contains the details leading to 
\begin{equation}
T^{L,(+)}_{lm}(k\hat z) = (i)^{l+1}\left[\delta_{m,2}-\delta_{m,-2}\right] \sqrt{2l+1} 
\frac{ \sqrt{\pi(l+2)(l+1)l(l-1)}}{2}k
\int_0^{D_s} dD \, T(k,\eta_0-D)  
\frac{j_l(kD)}{(kD)^2}
.\eql{tranlen}
\end{equation}
Fig.~\rf{tlww} shows the transfer functions for the lowest multipoles, assuming all galaxies are at redshift $z_s=1$. The amplitudes here are important: the transfer function for $l=2$ is largest, reflecting the fact that the signal will be largest on the largest scales.

\Sfig{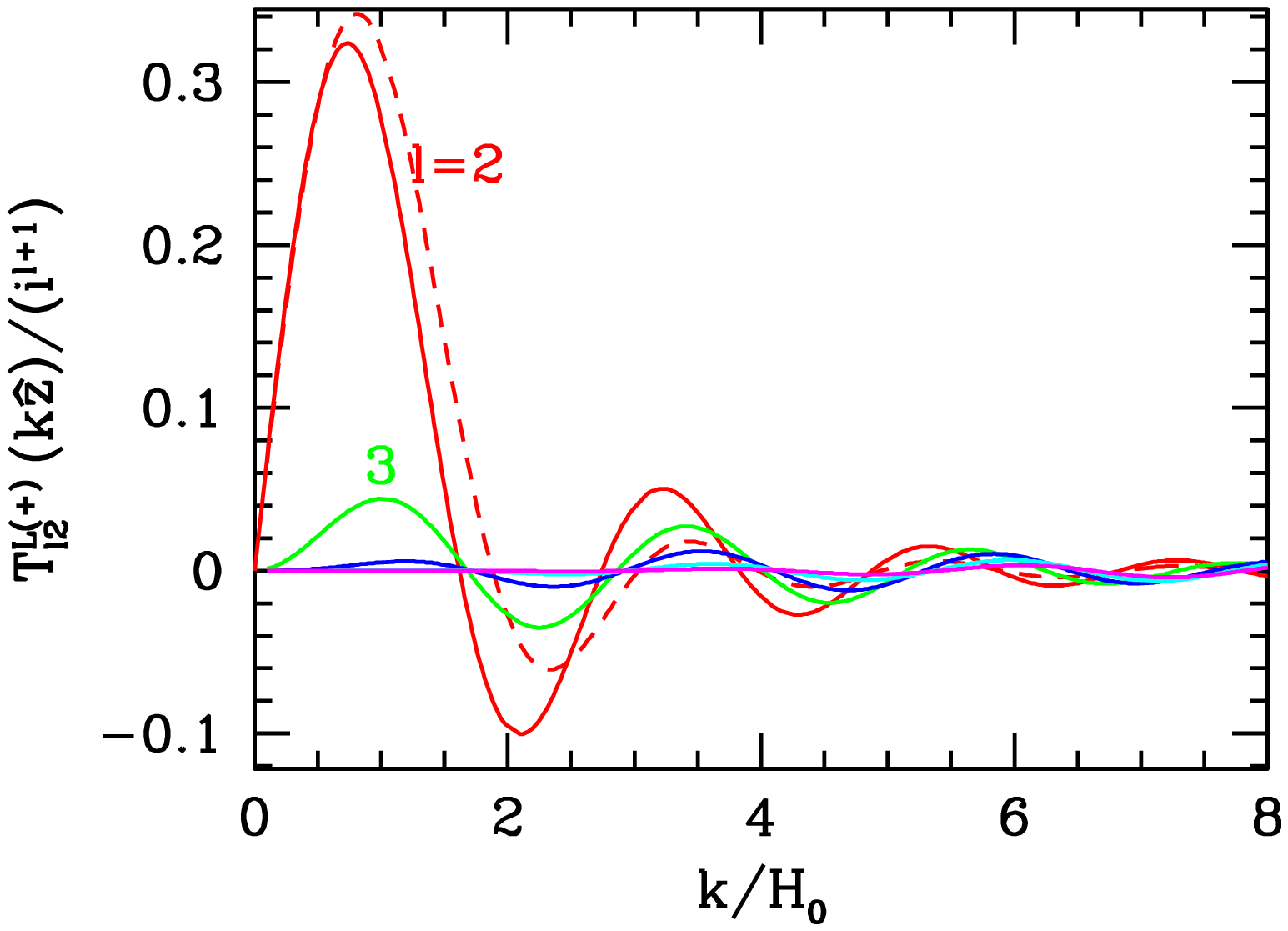}{The $m=2$ moment of the B-mode lensing transfer function for $l=2-6$. 
Sources are all assumed to be at $z=1$.}
%\appendix

\section{Cross-Correlation}

We can now collect the window functions for the two auto-spectra and the cross-spectrum and compute the correlation co-efficient. Explicitly, in addition to \ec{clbfin}, we have
\begin{eqnarray}
W^{LL}_l(k) &=& \pi(2l+1) 
(l+2)(l+1)l(l-1)) 
 \bigg\vert k
\int_0^{D_s} dD \, T(k,\eta_0-D)  
\frac{j_l(kD)}{(kD)^2}\bigg\vert^2\vs 
W^{PL}_l(k) &=& 6\pi 
\sqrt{(l+2)(l+1)l(l-1)}k
\int_0^{D_s} dD' \, T(k,\eta_0-D')  
\frac{j_l(kD')}{(kD')^2}\int_{0}^{D_{\rm reion}} dD\, 
\vs 
&&\times
 \dot\tau\left(\eta_0-D\right) \left[ (l+2) j_{l-1}(kD) -(l-1) j_{l+1}(kD) \right] 
 \int_{0}^{\Delta D} d(\Delta D') \, \dot T(k,\eta')
\frac{j_2(k\Delta D')}{(k\Delta D')^2}
\end{eqnarray}

The correlation coefficients which express the degree to which the moments are correlated 
are defined as
\begin{equation}
\alpha_l \equiv \frac{ C^{PL}_l}{\sqrt{C^{PP}_l C^{LL}_l} }
\end{equation}
where each $C_l$ is an integral over the power spectrum modulated by the window function $W_l$. 
The window functions for the lowest moments of the cross-spectra are plotted in Fig.~\rf{wlcross}
when the background galaxies in the lensing survey are all assumed to be at $z=1$ and reionization takes place instantaneously at $z=1$. 
As is clear, the $l=2$ moment has a negative correlation coefficient and is particularly 
sensitive to modes of precisely the size of the horizon. The lowest correlation coefficients -- again
when $z_{\rm source}=1$ and $z_{\rm reion}=10$ --  
are: $(\alpha_2,\alpha_3,\alpha_4,\alpha_5,\alpha_6=-0.32,0.10,0.31,-0.09,-0.20)$.

\Sfig{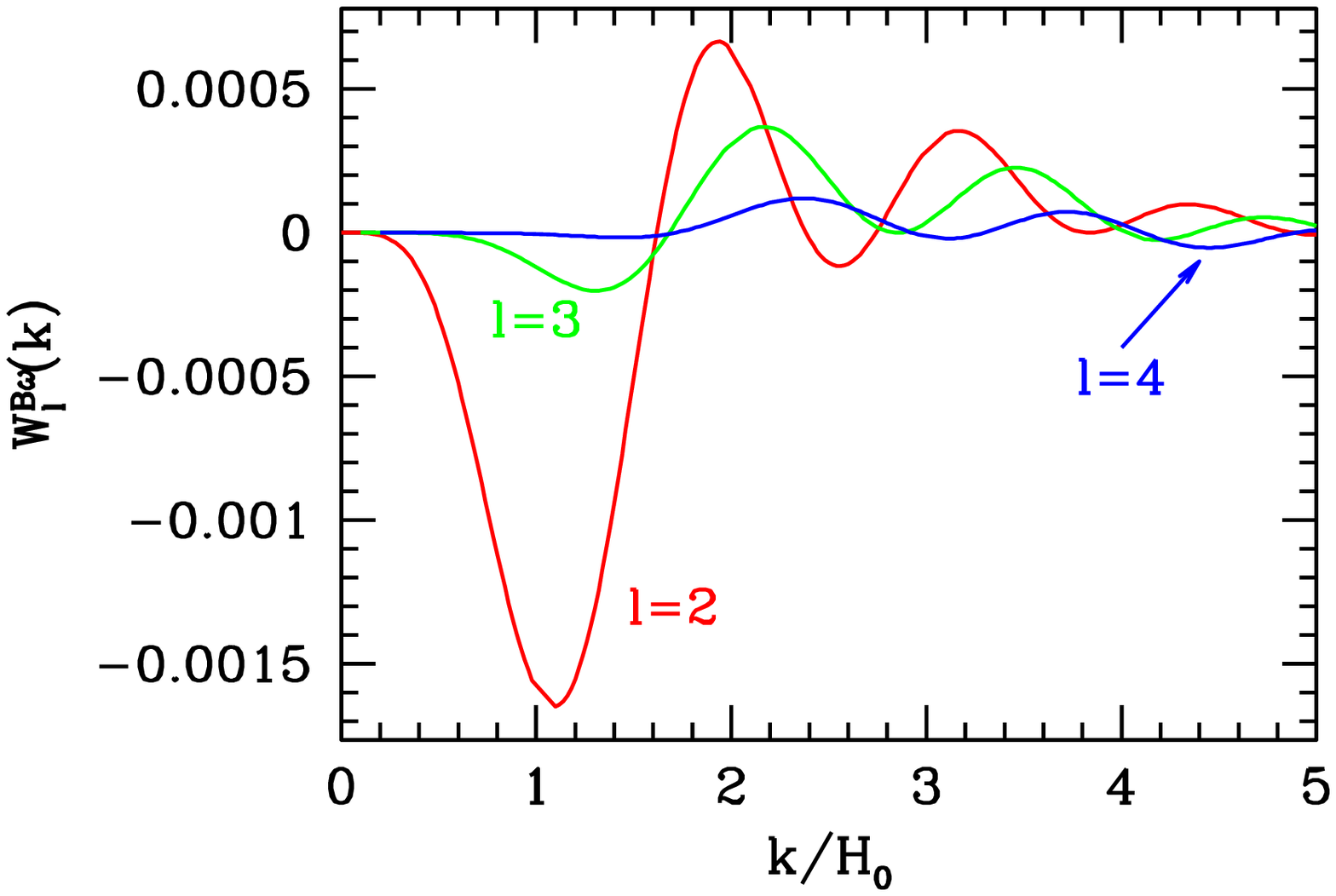}{Cross-correlation window function for lowest moments when lensing galaxies are all at $z=1$ and reionization occurs instantaneously at $z_{\rm reion}=10$.}

A fascinating aspect of the cross-correlation is that, because of the oscillations in the
gravitational waves, the correlation coefficient is sensitive to the redshift of the source
galaxies in the lensing survey. Fig.~\rf{zs} shows this dependence. In principle, one could imagine exploiting this dependence by
weighting each galaxy by some factor to maximize the signal to noise extraction, but I do not pursue this possibility here.

\Sfig{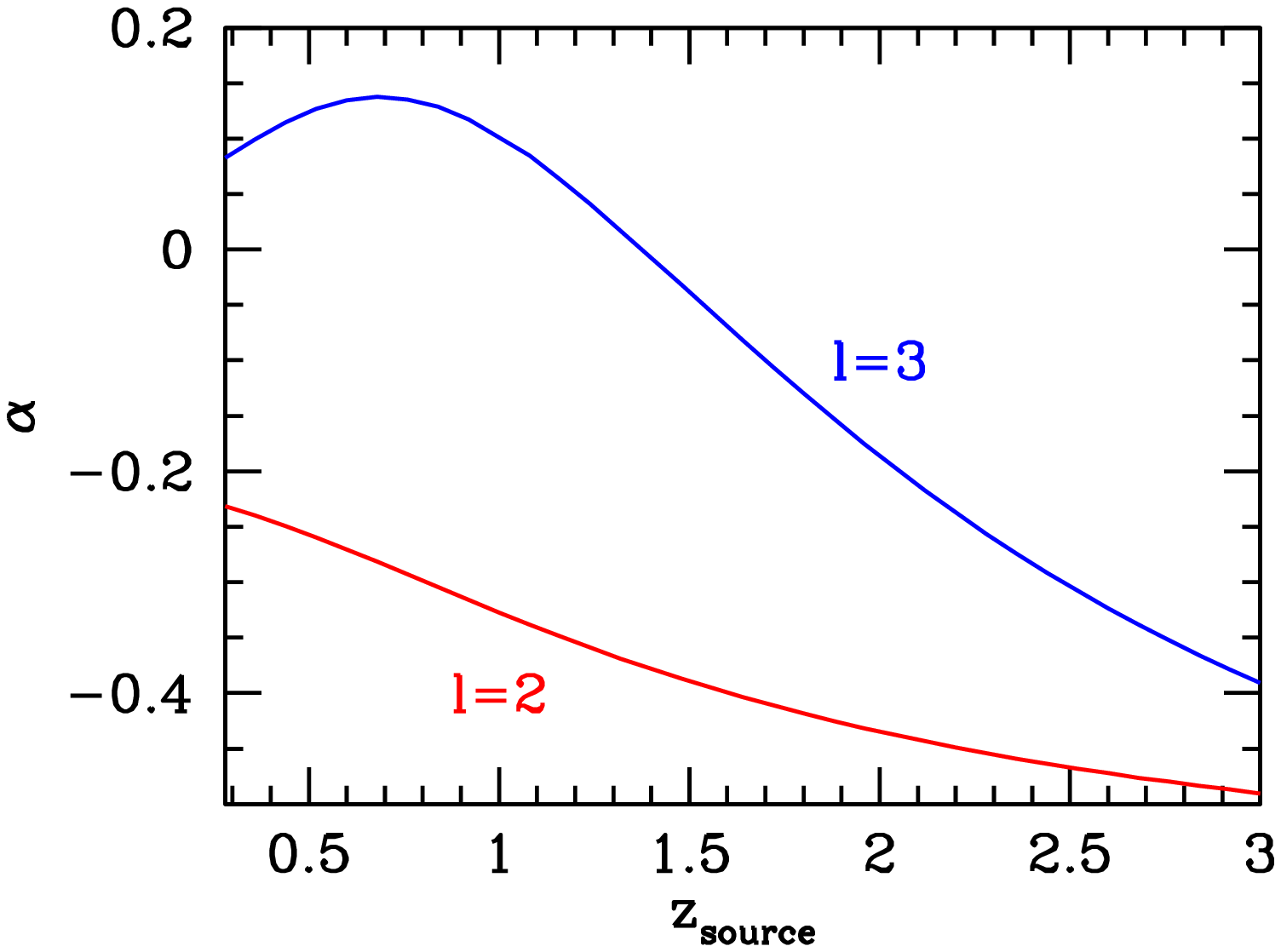}{Cross-correlation coefficient of CMB and weak lensing B-modes for the lowest two moments as a function of the redshift
of the galaxies in the lensing survey.}

\section{Significance}

How much would a lensing survey sensitive to the lowest moments add to our knowledge about the primordial gravitational waves?  And how does the cross-correlation affect this answer? As political pollsters have discovered,
the answers to these questions depend sensitively on how they are worded. 

For example, a simple question one might ask is how much tighter the constraints on the gravitational wave amplitude would become if the information from an all-sky lensing survey were added to that from a CMB polarization experiment. A quantitative answer to this can be obtained with the Fisher formalism. 
Consider two data points, the $l=2,m=2$ moment of the CMB B-mode and the same moment of the lensing B-mode. Normalize each so that the expected contributed to the variance from noise is unity. Also, allow for one free parameter, the normalization of the GW amplitude, $A$, with ``true'' value equal to 1. Finally, call the ratio of the signal variance to the noise variance $\lambda$ for each probe. Then the $2\times2$ covariance matrix is
\begin{equation}
C = \left( \matrix{ A\lambda_P + 1 & \alpha A\sqrt{\lambda_P\lambda_L}\cr
\alpha A\sqrt{\lambda_P\lambda_L} & A\lambda_L+1}
\right).
\end{equation}
Here $\alpha$ is the correlation coefficient computed in the previous section. The $_{11}$ entry contains contributions from both signal ($a\lambda_P$) and noise (1). The signal contribution to the variance is $\lambda_P\equiv C^{PP}_{l=2}/N^{PP}_{l=2}$ where $C^{PP}$ is the power computed in \S{II} and $N^{PP}$ is the noise variance which of course depends on the experiment. Here and throughout this section, I will focus only on the lowest $l=2$ moment since lensing contributions fall off rapidly with $l$. For orientation, current bounds on the gravitational wave amplitude from the CMB restrict $C^{PP}_{l=2}$ to be of order $(100\,{\rm nK})^2$ while future experiments~\cite{Bock:2009xw} aim for noise as low as $N^{PP}\sim (1 \,{\rm nK})^2$, so thoughts of $\lambda_P$ as large as $\sim 10^4$ are not crazy. Our goal is to estimate the uncertainty on the one free parameter $A$. The off-diagonal elements are free from noise so depend only on the cross-correlation of the two signals. Starting from this covariance matrix, and including all five $l=2$ moments, the standard formula for the Fisher matrix from a Gaussian process leads to a fractional error on $A$ of 
\begin{equation}
\frac{\Delta A}{A} = \sqrt{\frac{2}{5}} \frac{1+\lambda_L + \lambda_P+ \lambda_P\lambda_L(1-\alpha^2)}
{\left[ 2\lambda_P^2\lambda_L^2(1-\alpha^2)^2 + 2(1-\alpha^2)\lambda_P\lambda_L(\lambda_P+\lambda_L) + \lambda_L^2+\lambda_P^2+2\alpha^2\lambda_P\lambda_L\right]^{1/2}}.
\end{equation}
The factor of $\sqrt{5}$ in the denominator here comes from summing all five $l=2$ harmonics.

In the limit that $\lambda_P$ is large (corresponding to high signal to noise detection in the CMB), Fig.~\rf{fisher} shows how the constraints on the amplitude depend on the correlation coefficient and the lensing signal to noise. When the lensing signal is very small ($\lambda_L=0.1$) the error is simply equal to $\sqrt{2/5}=0.63$, the minimum possible error due to the cosmic variance of the $l=2$ CMB mode. If the signal to noise of lensing is higher, then the fractional error can go down, in principle as low as $\sqrt{1/5}$ since the amount of information doubles. Fig.~\rf{fisher} illustrates that correlations {\it degrade} the extraction of the amplitude. This makes sense: when trying to measure a variance one wants
	as many independent numbers as possible. If $\alpha=1$, then the two
	numbers (B-modes from CMB polarization and from lensing) are not independent 
	so the error on the variance increases. As $\alpha\rightarrow 1$, the
	constraint on $A$ reverts back to the $\sqrt{2/5}$ limit that would be obtained without the additional lensing information.

The conclusion from this exercise is that lensing information would help very little in the effort to pin down the gravitational wave amplitude. At best -- information only from $l=2$ and large signal to noise from both sets of experiments -- the reduction in the error would be only a factor of $\sqrt{2}$. Most likely, if GW are detected, the $l>2$ moments in the CMB will be very important while those higher moments will be undetectable in lensing. So the gain from a lensing survey would be diluted significantly. The effect studied here -- cross-correlations between the two signals -- serves to further dilute the impact of lensing, as the information would be redundant and hence useless in constraining the GW amplitude. However, the results of the previous section suggest that $\alpha^2$ is likely to be small enough so that the dilution would be minimal. If the lensing signal were detected, it would provide -- for the most part -- independent information about the GW amplitude.

\Sfig{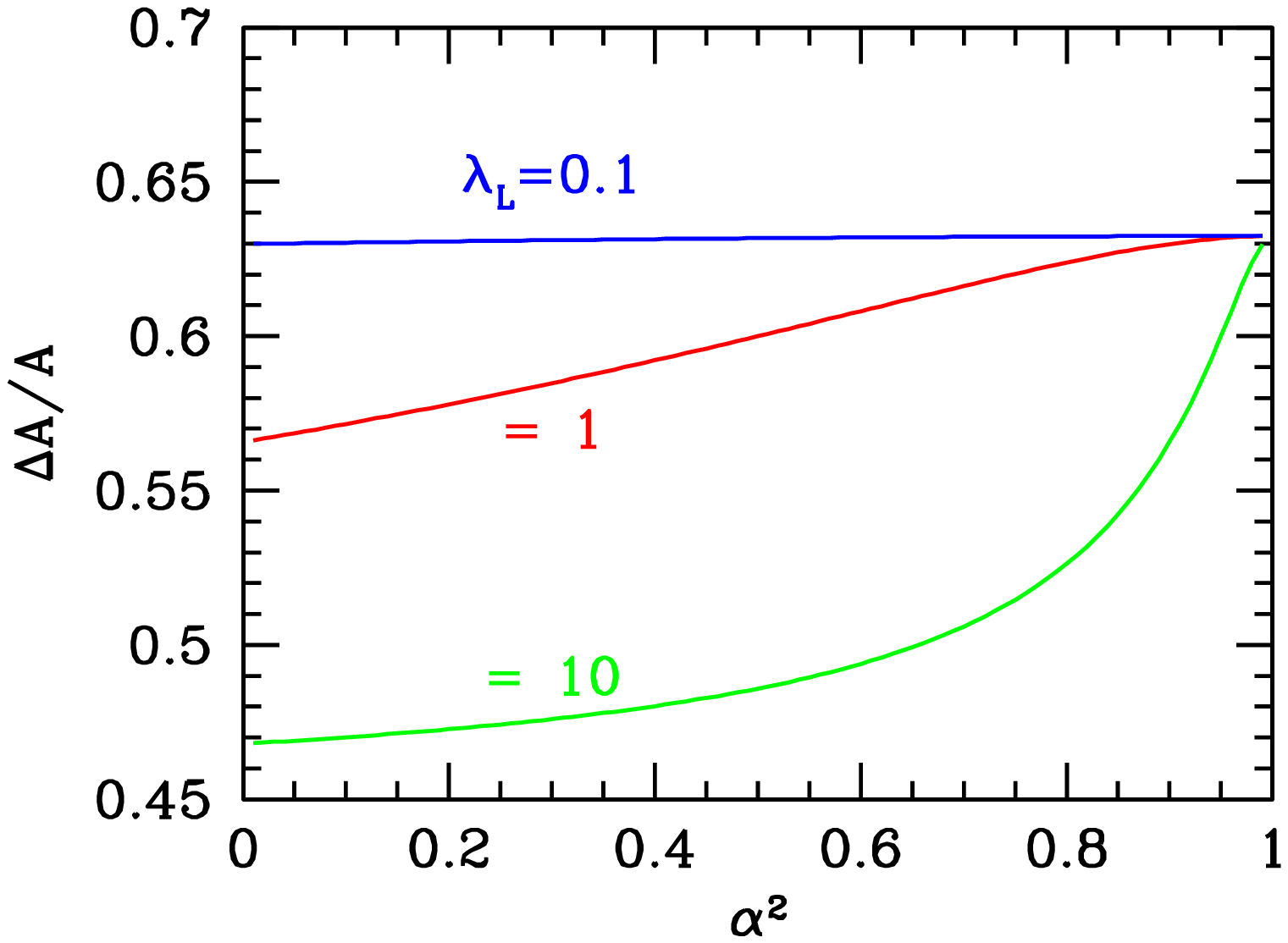}{Constraints on the gravitational wave amplitude as a function of
the cross-correlation co-efficient if the signal in the CMB experiment is large. The signal to noise in the lensing survey is
$\lambda_L$. For large values of $\lambda_L$, the error on the gravitational wave amplitude goes down by $\sqrt{2}$ unless correlations
between the two sets of measurements are large ($\alpha\rightarrow 1$).}

% WMAP detection of anisotropies from l=2. Detected is 2      213.4069      934.2714        5.5978
% So \lambda = 934/5.6
% L = exp(-213/2(934+1)}/934^{5/2}
% while L(C=0)
% 

Another way to probe the importance of lensing and cross-correlations is to focus not on parameter determination ($\Delta A$) but rather on the firming up the case for detection. To show the difference between these two sets of questions, consider 
a simple example: temperature anisotropies in the $l=2$ mode as measured by WMAP~\cite{wmap}. The expected value of $l(l+1)C_l/(2\pi)=3C_2/\pi$ in the standard $\Lambda$CDM model is $928 (\mu K)^2$, while the measurement error contributes only $0.01 (\mu K)^2$, so the value of $\lambda$ is this case [the ratio of cosmic variance to measurement error] is equal to $928/0.0124=75,000$. The Fisher one-sigma error on the fractional amplitude is $\sqrt{2/5}(1+1/\lambda)$, essentially equal to the 0.71 cosmic variance limit. This large error hides the extreme non-gaussianity of the likelihood function. Indeed, even using $l=2$ only, the statistical probability that there is no signal is infinitesimally small. To quantify this, consider the ratio of the likelihoods for two different models: (i) the best fit $\Lambda$CDM model with $2C_2/\pi\simeq 928 (\mu K)^2$ and (ii) a model where $C_2=0$ with no signal. Since the measured value of $3C_2/\pi$ is $201 (\mu K)^2$, the ratio of these two likelihoods -- using only $l=2$ data -- is
\begin{equation}
\frac{\mathcal{L}_1}{\mathcal{L}_2} = \frac{1}{(1+\lambda)^{5/2}} \exp\left\{ -\frac{5}{2} \Big[
\frac{201/0.0124}{1+\lambda} - \frac{201}{0.0124} \Big]\right\}
\end{equation}
of order $e^{40,000}$! Therefore, while the measured $l=2$ moments give very little information about the amplitude of the anisotropies, they weigh in very heavily on the question of whether or not anisotropies have been detected.

%
%
%
%
%
%
%only the CMB and focus only
%on the $l=2$ multipoles. The above suggests that using the lowest moment, one would never be able to detect gravitational waves in the CMB since the 1-sigma error on the amplitude will always be at least $\Delta A/A=\sqrt{2/5}$. If the likelihood were Gaussian, then no detection ($A=0$) could never be ruled out by more than $1.5\sigma$ with only the $l=2$ information.  However, the likelihood is far from Gaussian, and the likelihood for $A=0$ will often be much smaller than that for $A=1$. 

Returning to B-mode detection from the CMB, instead of using the one-sigma Fisher error, we compute the ratio of the likelihood for detection $\mathcal{L}(A=1)$ vs. the likelihood of no detection $\mathcal{L}(A=0)$. This likelihood ratio will vary depending on the data. Fig.~\rf{nofg} shows the distribution of the likelihood ratio $\mathcal{L}(A=1)/\mathcal{L}(A=0)$ for 1000 mock ``skies'' (here sky simply means the five $l=2$ CMB moments) generated from a true model with $A=1$.  Even when the signal to noise is unity ($\lambda_P=1$) so that the 1-sigma Fisher error is $\Delta A/A=1.26$, the likelihood ratio in a given experiment could be very large, signaling a detection. Ten percent of the mocks produced likelihood ratios greater than 100; that is, one would have concluded with 99\% certainty that there are B-modes. A 90\% CL
detection would have occurred 41\% of the time. If the signal to noise were larger, $\lambda_P=10$, then detection is virtually assured, with a 99\% detection emerging from 95\% of the runs.

\Sfig{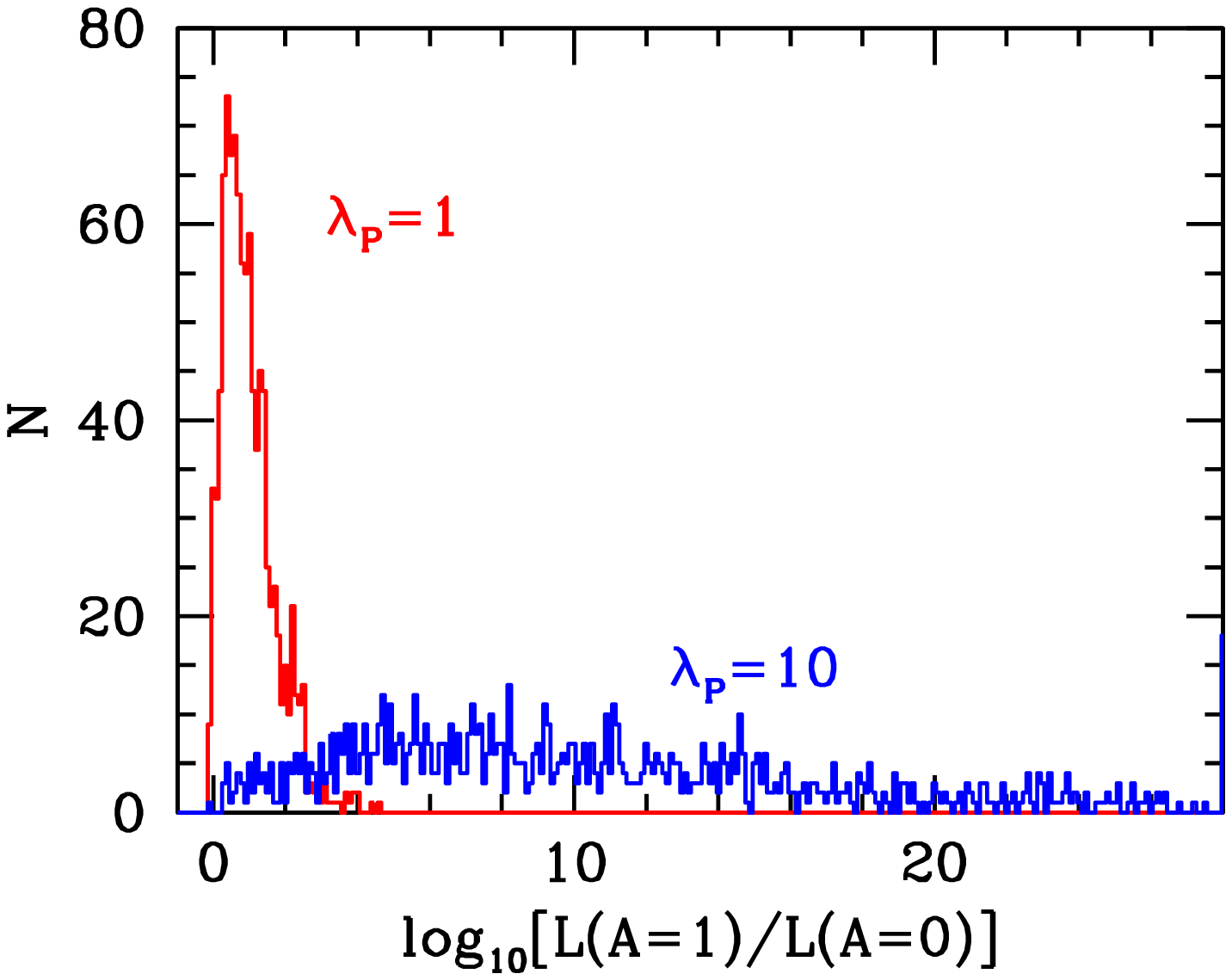}{Distribution of likelihood ratios for gravitational waves (GW) and no GW 
for 1000 fake skies generated when the true model contains GW's (using $l=2$ CMB data only). When the underlying model has $\lambda_P=1$,
the Fisher estimated 1-sigma fractional error on the amplitude of GW's would be $\sqrt{2/5}\times (1+\lambda_P)/\lambda_P = 1.26$, while the $\lambda_P=10$ model produces a Fisher error of $0.7$. Nonetheless, a statistically significant detection is possible in the first case and virtually assured in the second.}

This suggests that, while lensing will not be of much use in constraining the GW amplitude, it might help in firming up the evidence for a detection. 
For example, suppose a CMB experiment measured a non-zero B-mode, but one wanted to compute the likelihood that this was due to GW or to foregrounds. With a CMB experiment only, the likelihood ratio of these two ``models'' (signal due to GW or signal due to foregrounds) would be unity: there would no way to tell them apart. How much would a lensing experiment improve on this? To answer this, I generated 1000 realizations of both lensing and CMB $l=2$ moments with $\lambda_P=10$ and $\lambda_L=1$ and a given value of $\alpha$. For each realization,
I computed the likelihood ratio of two models:
\begin{itemize}
  \item{\bf Model 1: True Model} Signal to Noise in CMB $\lambda_P=10$; in lensing $\lambda_L=1$ and the true value of $\alpha$ 
  \item{\bf Model 2: Null Model} Foregrounds produce $\lambda_P=10, so \lambda_L=0$ 
  \end{itemize}
  The distribution of these likelihood ratios is shown in Fig.~\rf{histo} for several values of $\alpha$. 
When $\alpha=0$, (7,39)\% of the realizations led to a (99,90)\% detection;
For $\alpha=0.5$ those percentages go up to (12,45), and for larger values of $\alpha=0.9$ up to
(21,57). So when the question posed is one of detection, as opposed to parameter determination, non-zero cross-correlation is beneficial.

\Sfig{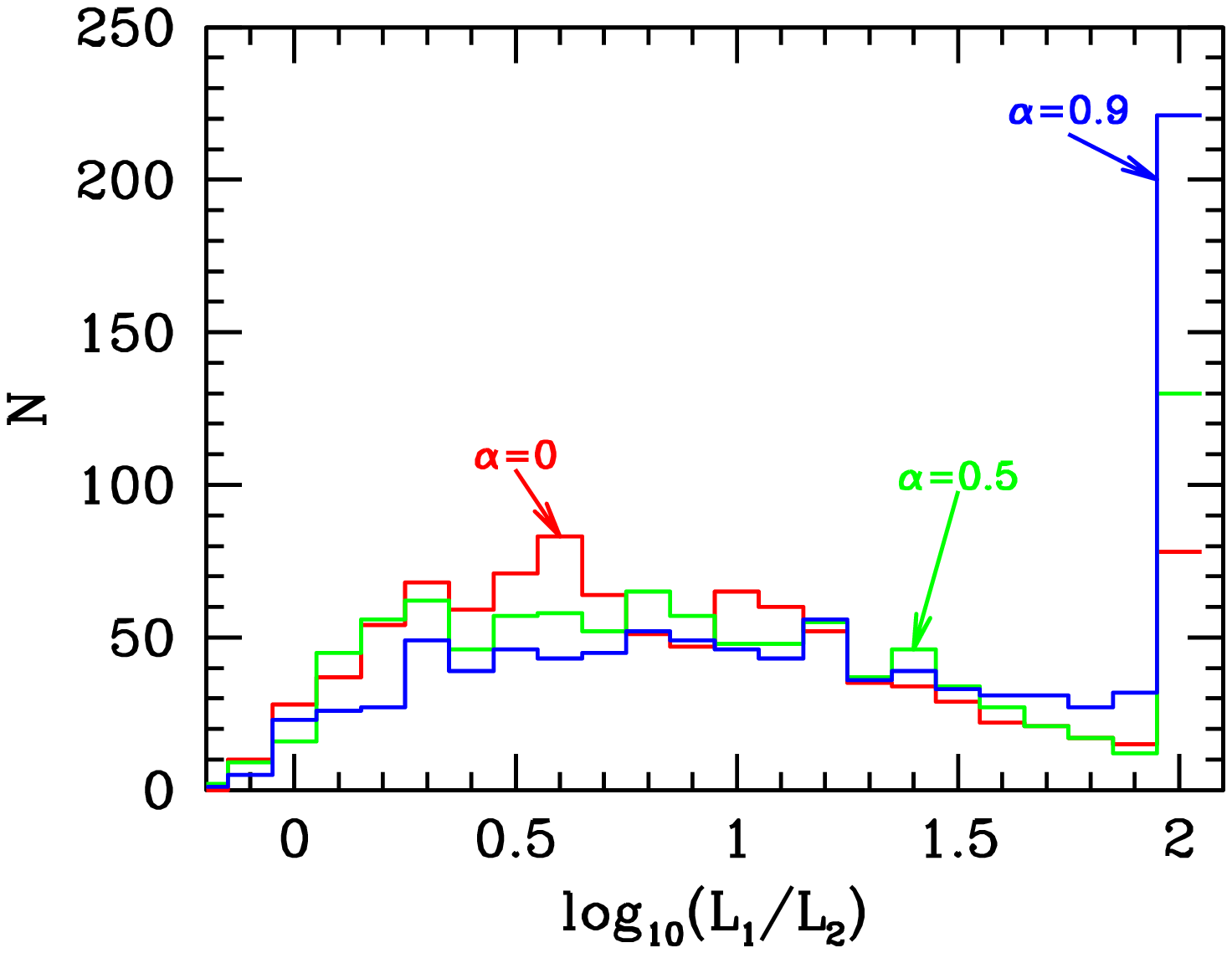}{Likelihood ratios of GW vs. (foregrounds in CMB and no signal in lensing) for 1000
mock skies generated from a true model with GW.}

A harder question is whether one could detect the cross-correlation. To approach this question, I set the true values of $\lambda_P=100$ and $\lambda_L=1$ and computed the likelihood ratio of two models with: 
\begin{itemize}
  \item{\bf Model 1:} $\alpha$ equal to its true value
  \item{\bf Model 2:} $\alpha=0$
  \end{itemize}
The likelihood ratio was rarely significant. A 90\% detection is never obtained if $\alpha=0.5$ and only 11\% of the time even if
$\alpha=0.95$. So we are unlikely to detect the cross-correlation of the lensing and CMB B-modes.

%\Sfig{corr}{Likeihood ratios for $\alpha=\alpha_{\rm true}$ vs. $\alpha=0$ for 1000 mocks.}

\section{conclusions}

Primordial gravitational waves produce indirect effects on both the polarization of the CMB and the lensing of distant galaxies. These effects are correlated, with a size and sign which depends on the angular scale and on the redshifts of the background galaxies. The correlation may help sort out systematics if a B-mode detection is made in the CMB. It seems likely that all-sky lensing surveys will be carried out for other purposes, so the search for the B-mode in that arena will cost nothing. It might even be used to motivate such surveys should the CMB B-modes be detected. 

I thank Anthony Challinor, Rob Crittenden, Wayne Hu, Lam Hui, Matthew Kleban, Eiichiro Komatsu, Hiranya Peiris, and Albert Stebbins for helpful conversations. This work was supported by the DOE at Fermilab and by NSF Grant AST-0908072.

\bibliography{cmbl}

\appendix

\section{Polarization Calculation}\label{app:pol}

Here I compute the transfer functions when the wavevector $\vec k$ lies along the $\hat z$ axis.
Consider first the integral over $\hat n'$ in \ec{defj} for the case of $\alpha=+$ (the $\times$ contribution is identical so we need compute only this component). Thus,
\begin{eqnarray}
\epsilon_{ij}^{(+)}(\hat z)\hat n'^i \hat n'^j
&=& (\hat n'^x)^2 - (\hat n'^y)^2
\vs 
&=&
 \sin^2\theta\cos(2\phi)\vs 
&=&
- \sqrt{\frac{8\pi}{15}} \left( Y_{22}(\hat n') + Y_{2-2}(\hat n')\right)
\end{eqnarray}
where $\theta$ and $\phi$ are the standard spherical angles with respect to the $z$-axis. Meanwhile the exponential can be expanded in spherical harmonics, or since $\vec k$ is taken along the $z$-axis:
\begin{equation}
e^{i\vec k\cdot \hat n' \Delta D'} = \sum_{l'=0}^\infty \sqrt{4\pi(2l'+1)} i^{l'} Y_{l'0}(\hat n') j_{l'}(k\Delta D').
\end{equation}
Therefore,
\begin{equation}
J^{(+)}_{m''}(k\hat z,\Delta D') = - \sqrt{\frac{32\pi^2}{15}}\sum_{l'=0}^\infty \sqrt{2l'+1} i^{l'}  j_{l'}(k\Delta D')
\int d^2n'\, Y^*_{2m''}(\hat n') Y_{l'0}(\hat n') \left( Y_{22}(\hat n') + Y_{2-2}(\hat n')\right).
\end{equation}
The integral of the product of the $Y_{lm}$'s over $\hat n'$ can be expressed in terms of the Wigner 3$j$-symbols:
\begin{equation}
\int d^2n'\, Y^*_{2m''}(\hat n') Y_{l'0}(\hat n') \left( Y_{22}(\hat n') + Y_{2-2}(\hat n')\right)
= \sqrt{\frac{25(2l'+1)}{4\pi}} \left( \matrix{2&2&l'\cr 0&0&0} \right)
\left[ \bigg( \matrix{2&2&l'\cr 2&m''&0} \bigg) + \bigg( \matrix{2&2&l'\cr -2&m''&0} \bigg) \right].
\end{equation}
The first 3$j$-symbol is non-zero only when $l'=0,2,4$, and then is
\begin{equation}
\left( \matrix{2&2&l'\cr 0&0&0} \right) 
= (-1)^{l'/2} \frac{l'!}{[(l'/2)!]^2} \left[ \frac{(4-l')!}{(5+l')!}\right]^{1/2}
\frac{(2+l'/2)!}{(2-l'/2)!} .
\ee
The first 3$j$-symbol in the square brackets is zero unless $m''=-2$ while the second is non-zero only if $m''=2$ and is equal to the first, so
\begin{equation}
J^{(+)}_{m''}(k\hat z,\Delta D') = - \sqrt{\frac{40\pi}{3}}\sum_{l'=0,2,4}(2l'+1)  j_{l'}(k\Delta D')
  \frac{l'!}{[(l'/2)!]^2} \left[ \frac{(4-l')!}{(5+l')!}\right]^{1/2}
\frac{(2+l'/2)!}{(2-l'/2)!}
\left[ \delta_{m'',-2} + \delta_{m'',2}\right]\bigg( \matrix{2&2&l'\cr 2&-2&0} \bigg) .
\end{equation}
The relevant 3$j$-symbols with $l'=0,2,4$ are
$1/\sqrt{5},\sqrt{2/35},$ and $1/\sqrt{630}$ respectively.
Therefore,
\begin{eqnarray}
J^{(+)}_{m''}(k\hat z,\Delta D') &=& -\sqrt{\frac{40\pi}{3}}  \left[ \delta_{m'',-2} + \delta_{m'',2}\right] \frac{ 7
j_{0}(k\Delta D') + 10 j_{2}(k\Delta D')+3  j_{4}(k\Delta D')}{35}
\vs &=&
- \sqrt{\frac{40\pi}{3}} \left[ \delta_{m'',-2} + \delta_{m'',2}\right] \frac{3j_2(k\Delta D')}{(k\Delta D')^2}
  .
\end{eqnarray}

A similar calculation for the integral in \ec{defi} begins by expanding the exponential in terms of spherical harmonics. We then are left with an integral over the product of terms of the form $Y_{l'm'}(\hat n)  {}_{\pm 2}Y_{2m''}(\hat n)  {}_{\pm 2}Y^*_{lm}(\hat n)$. The first two of these can be re-expressed as a sum over ${}_{\pm 2}Y_{LM}$ using Eq. (8) from Ref.~\cite{Hu:1997hp}. The integral will pick out only the term with $L=l$ and $M=m$, so
in terms of the Clebsch-Gordon coefficients,
\begin{equation}
Y_{l'm'}(\hat n)  {}_{\pm 2}Y_{2m''}(\hat n) \rightarrow
\left[ \frac{5 (2l'+1)}{4\pi (2l+1)}\right]^{1/2} 
\langle 2,l';m'',m'\vert 2,l';l,m\rangle
\langle 2,l';\mp2,0\vert 2,l';l,\mp 2\rangle {}_{\pm 2}Y_{lm}(\hat n).
\end{equation}
The integral is then
\begin{equation}
I_{lmm''}(k\hat z,D) = \sqrt{5} \delta_{m,m''} \sum_{l'} (2l'+1) (-i)^{l'} j_{l'}(kD) 
\langle 2,l';m,0\vert 2,l';l,m\rangle
\left[ \bigg(\matrix{l'&2&l\cr0&-2&2}\bigg) - \bigg(\matrix{l'&2&l\cr0&2&-2}\bigg)\right]
\end{equation}
where one of the Clebsch-Gordon sets has been written in terms of Wigner $3j$-symbols and the identity $Y_{l,m-m''}(\hat z)= (-1)^l\delta_{m,m''}\sqrt{2l+1}/\sqrt{4\pi}$ has been used. The second Wigner $3j$-symbol is identical to the first apart from a sign flip in the bottom row, which changes the overall sign by $(-1)^{l'+2+l}$. Only if that sign is negative will cancellation be avoided, so one requirement is that $l'+l$ is odd. The values of $l,l'$ for which these $3j$-symbols are non-zero are those which satisfy: $|l-l'|\le 2\le |l+l'|$. The only values of $l'$ which satisfy both conditions are $l'=l\pm 1$. Therefore,
\begin{eqnarray}
I_{lmm''}(k\hat z,D) &=& 2 \sqrt{5} \delta_{m,m''} \Bigg[
(2l-1) (-i)^{l-1} j_{l-1}(kD) 
\langle 2,l-1;m,0\vert 2,l-1;l,m\rangle
\bigg(\matrix{l-1&2&l\cr0&-2&2}\bigg) \vs 
&&+
(2l+1) (-i)^{l+1} j_{l+1}(kD) 
\langle 2,l+1;m,0\vert 2,l+1;l,m\rangle
\bigg(\matrix{l+1&2&l\cr0&-2&2}\bigg) 
\Bigg].
\end{eqnarray}
Since we will be multiplying this by $J^{(\alpha)}_{m''}$ which above was shown (in the case $\hat k\parallel\hat z$) to vanish unless $m''=\pm2$, 
we need evaluate the Clebsch-Gordon coefficients only for $m=m''=\pm2$. For example,
\begin{eqnarray}
\langle 2,l-1;2,0\vert 2,l-1;l,2\rangle \bigg(\matrix{l-1&2&l\cr0&-2&2}\bigg)
&=& (-1)^{l}\sqrt{2l+1}\bigg(\matrix{l-1&2&l\cr0&2&-2}\bigg)\bigg(\matrix{l-1&2&l\cr0&-2&2}\bigg)
\vs 
&=&  4 (-1)^{l+1}\sqrt{2l+1} \frac{(2l-3)! (l+2)!}{(2l+2)!(l-2)!} 
\end{eqnarray}
so the coefficient of $j_{l-1}$ for $m=2$ is
\begin{equation}
8 \sqrt{5}  (2l-1) (-i)^{l-1} (-1)^{l+1}\sqrt{2l+1}\frac{(2l-3)! (l+2)!}{(2l+2)!(l-2)!}  
=i^{l-1} \sqrt{\frac{5}{2l+1}} (l+2).
\end{equation}
Using the symmetries of the $3j$-symbols, it is easy to see that the $m=-2$ term is equal and opposite. Carrying out similar steps for the coefficient of $j_{l+1}$ leads finally to
\begin{equation}
I_{lmm''}(k\hat z,D) = i^l \sqrt{\frac{5}{2l+1}} \delta_{m,m''} \left(\delta_{m,2}-\delta_{m,-2}\right)
\left[ (l+2) j_{l-1}(kD) -(l-1) j_{l+1}(kD) \right].
\end{equation}
Therefore the transfer function in \ec{tran} reduces to \ec{tranfin}.

\section{Weak lensing Calculation}\label{applen}

In this case, contracting the indices and expanding the exponential leads to
\begin{equation}
T^{L,(+)}_{lm}(k\hat z) = k\int_0^{D_s} dD \, T(k,\eta_0-D) \sum_{l'=0}^\infty i^{l'-1} \sqrt{ \frac{(2l'+1) 8\pi^2}{15}}
j_{l'}(kD)
\int d^2n Y^*_{lm}(\hat n) Y_{l'0}(\hat n) \left[ Y_{22}(\hat n) - Y_{2-2}(\hat n) \right].
\end{equation}
The integral over the product of spherical harmonics can again be expressed in terms of the Wigner $3j-$symbols, leading to
\begin{eqnarray}
T^{L,(+)}_{lm}(k\hat z) &=& \sqrt{ \frac{2 (2l+1) \pi}{3}} k
\int_0^{D_s} dD \, T(k,\eta_0-D) \sum_{l'=0}^\infty i^{l'-1} (2l'+1)
j_{l'}(kD)\vs 
&&\times 
\left( \matrix{l&l'&2\cr0&0&0}\right) \bigg[ \left( \matrix{l&l'&2\cr-m&0&2}\right) 
- \left( \matrix{l&l'&2\cr-m&0&-2}\right)  \bigg].
\end{eqnarray}
Only $l'=l,l\pm2$ are non-zero here, so the integrand will have terms proportional to $j_l,j_{l\pm2}$. The coefficients of these terms, however, are such that the resulting sum is proportional to $j_l(kD)/(kD)^2$, so
\begin{equation}
T^{L,(+)}_{lm}(k\hat z) = (i)^{l-1}\left[\delta_{m,-2}-\delta_{m,2}\right] \sqrt{ \frac{2 (2l+1) \pi}{3}} k
\int_0^{D_s} dD \, T(k,\eta_0-D) \frac{\sqrt{6}}{2} \frac{ \sqrt{(l+2)(l+1)l(l-1)}}{2}\frac{j_l(kD)}{(kD)^2}
.
\end{equation}
This leads directly to \ec{tranlen}.

\end{document}